# Theoretical analysis of supercontinuum and colored conical emission produced during ultrashort laser pulse interaction with gases

V. V. Semak[1] and M. N. Shneider[2]

[1] Applied Research Laboratory, the Pennsylvania State University, PA
[2] Department of Mechanical and Aerospace Engineering, Princeton University, NJ

**Abstract**
We use a conceptually new approach to theoretical modeling of self-focusing in which we integrated diffractive and geometrical optics in order to explain and predict emission of white light and colored rings observed in ultrashort laser pulse interaction. In our approach laser beam propagation is described by blending solution of linear Maxwell's equation and a correction term that represents nonlinear field perturbation expressed in terms of paraxial ray-optics (eikonal) equation. No attempt is made to create appearance of exhaustive treatment via use of complex mathematical models. Rather, emphasis is placed on elegance of the formulations leading to fundamental understanding of underlying physics and, eventually, to an accurate practical numerical model capable of simulating white light generation and conical emission of colored rings produced around the filament.

**Introduction**

In our previous publication [1] we proposed a new approach to theoretical analysis of nonlinear refraction of a laser beam in gaseous medium due to Kerr effect and multiphoton gas ionization. In our approach laser beam propagation is described by combination of linear Maxwell's equation solution and a correction term that represents nonlinear wave front perturbation expressed in terms of paraxial ray-optics (eikonal) equation. Such approach is based on the assumption that perturbations of the radial profile of laser beam intensity and temporal profile of laser pulse can be neglected while focused laser beam propagates through the near focal zone of high intensity (approximately several Rayleigh lengths). The results of our computations confirm validity of the assumptions of negligible perturbation of temporal and radial profiles of the beam within the computational zone for the most of the real-life applications of modern ultrashort pulse lasers. Additionally, in the description of beam propagation we neglected group velocity dispersion, self-steepening, and absorption and we assumed that Kerr effect is instantaneous (i.e., we neglect all effects associated with nonzero relaxation time).

In contrast to the alternative theoretical approaches that attempt description of nonlinear laser beam propagation in terms of complex mathematical models, such as, for example, self-similar solution of nonlinear Shroedenger's equation [2,3], our method delivers clear picture of the underlying physics and provides valuable and functional information. Previously our theoretical approach [1] demonstrated existence and quantified power losses from the periphery of the beam that reduce amount of laser power participating in self-induced refraction. Additionally, our approach showed absence of spatial and temporal balance between self-focusing and plasma defocusing of the laser beam and, thus, exposed shortcoming of the concept of "critical power" and futility of attempts to describe laser beam propagation in nonlinear media as a self-similar solution of a propagation equation.

In the present work we extended application of our theoretical approach to explain and simulate the laser induced white light (referred to sometimes as supercontinuum) and colored conical emission. The supercontinuum generation was first reported at the end of past century by Alfano and Shapiro [4,5] who observed laser spectrum broadening during picosecond laser interaction with different crystals and glass. Recent progress in ultrashort pulse laser technology revived broad interest to this effect and generated research effort [6] that, in particular, included studies of generation of white light and conical emission during laser interaction in gases. The experiments showed that if a laser beam with pulse duration lesser than approximately 1 ps and with sufficiently high energy is focused in air, at some (typically large ~ 10 meters) distance beyond the focus the beam cross section consists of a white central part surrounded by a rainbow-like conical emission (see, for example, [7] reporting white core surrounded by Newton's rings having a divergence of the order of the mrad or [8] reporting conical emission measured at 10 -100 m distance from the exit window of a 5 mJ, 45 fs, 800 nm pulse). It is typically observed that the radial order of the spectral components is such that the higher frequencies are at the larger radial distances [9].

Several models were proposed for "this still not entirely understood phenomenon" [9]: Cherenkov radiation [7,10], four-wave mixing [11, 12], self-phase modulation [13], and X-waves [14-16]. In our model we adopted self-phase modulation as a mechanism for the spectral broadening and the equation containing terms for the frequency shift due to medium nonlinearity and gas ionization was integrated into our mathematical model for propagation [1].

**Theoretical model**
*The propagation equation*
The propagation of a laser beam in a media with uniform and laser irradiance independent refractive index was described by Kogelnik and Li [17] who found approximate (corresponding to the paraxial propagation) solution of the wave equation. This solution is expressed via complex beam parameter, $q(z)$ which depends on the coordinate, $z$, along the beam axis and describes the Gaussian variation in beam irradiance with the distance, $r$, from the optic axis, as well as the curvature of the phase front, $R(z)$, which is spherical near the axis.

The complex beam parameter is given by the equation:

$$\frac{1}{q(z)} = \frac{1}{R(z)} - i\frac{\lambda}{\pi w(z)}, \tag{1}$$

where $\lambda$ is the wavelength and $w(z)$ is the measure of beam irradiance decrease by factor of $1/e^2$ with the distance from the axis. The parameter $w$ is often called the beam radius or "spot size".

For a focused laser beam according to the convention, the $z$ axis coordinate is set to zero at the beam waist and for a beam propagating from left to right; the $z$ values are negative to the left and positive to the right from the waist location. The wave front curvature radius is given by the equation [17]

$$R(z) = z\left(1 + \left(\frac{\pi w_0^2}{\lambda z}\right)^2\right), \tag{2}$$

where, $w_0$ is the beam radius at the $1/e^2$ level of the maximum beam irradiance in the beam waist (focal plane) and $\lambda$ is the laser wavelength. Thus, the diffraction dominated beam is converging for coordinates $z<0$ when the radius of the wave front curvature is negative, and it is diverging for $z>0$ when the radius of the wave front curvature is positive. In the beam waist ($z=0$) the beam front is plane.

For diffraction dominated beam propagation, and under assumption of paraxial propagation, the angle of an individual ray can be defined as arcsine of the ratio of radial position $r$ and the local wave front curvature radius, $R(z)$:

$$\varphi^D(r,z) = \arcsin\left(\frac{r}{R(z)}\right). \tag{3}$$

The laser beam radius, $w(z)$, at the intensity level of $1/e^2$ at the location along the beam axis with coordinate $z$ that is given by the equation

$$w^2(z) = w_0^2\left[1+\left(\frac{\lambda z}{\pi w_0^2}\right)^2\right], \tag{4}$$

and, the beam irradiance is

$$I(r,z,t) = \frac{2P(t)}{\pi w^2(z)}\exp\left(-\frac{2r^2}{w^2(z)}\right), \tag{5}$$

where $P(t)$ is time dependent (instantaneous) laser power.

Within geometrical optics interpretation, for a laser beam propagating length $ds$ along its trajectory in a medium with variable (both in time and space) refractive index the angle of an individual ray (the angle between the wave vector and the beam axis) $\varphi = \arcsin\left(\frac{dr}{ds}\right)$ can be derived from the eikonal equation [18]

$$\frac{d}{ds}\left(\frac{d\vec{r}}{ds}n\right) = \nabla n, \tag{6}$$

where $\vec{r}$ is the radial vector, $n$ is the refractive index, $dr$ is change of the radial vector along the coordinate normal to coordinate z that coincides with the beam axis. By convention, the positive value of the angle of an individual ray corresponds to divergence (the wave vector is directed away from the beam axis) and the negative value corresponds to convergence (the wave vector is directed toward the beam axis). For paraxial approximation that assumes small $\varphi$ and, therefore, $ds = dz/\cos\varphi \approx dz$, it follows from the equation (6) that

$$d\varphi \approx \frac{\partial n/\partial r}{n} dz, \tag{7}$$

where $d\varphi$ is the change of the angle of individual ray, $z$ is the coordinate along the beam axis, $r$ is the radial coordinate normal to the beam axis.

If a laser beam propagates in a media with nonlinear response and beam intensity is sufficiently high to produce media ionization, then in the reference coordinate system moving with the laser pulse ($\eta = t - \frac{nz}{c}$) the angle accumulated by an individual ray propagating from the location with coordinate $z_0$ to the location with coordinate $z_s$ can be expressed by the equation

$$\varphi^{NLP}(r, z_s, \eta) = \int_{z_0}^{z_s} d\varphi^{NLP} = \int_{z_0}^{z_s} \frac{\partial n/\partial r}{n} dz, \tag{8}$$

where refractive index, $n$, is function of laser beam irradiance, $I(r,z,\eta)$. The expression for the refractive index, $n$, is as follows

$$n = n_0 + n_2 I(r, z, \eta) - \omega_p^2 / 2\omega_l^2, \tag{9}$$

where $n_0$ is the linear index of refraction in air; $n_2$ is the second order nonlinear refractive index that equals the multiple of laser wavelength and Kerr constant; $\omega_p$ is the plasma frequency, $\omega_p^2 = \frac{e^2 N_e}{\varepsilon_0 m_e}$; $N_e$ is the density of electrons produced due to ionization; $m_e$ and $e$ are the mass and charge of electron, respectfully; and $\omega_l$ is the laser angular frequency.

It is easy to see that for the interaction of typical ultrashort pulse lasers wavelengths and pulse energies with air, the ultrashort laser pulses produce multiphoton ionization ($\gamma>1$, see [19]) with the electron number density given by the equation

$$N_e(r, z, t) = \sigma_k (N_0 - N_e) \int_{-\infty}^{\eta} [I(r, z, \xi)]^k d\xi, \tag{10}$$

where $N_0$ is the number density of neutrals, $\sigma_k$ is the ionization rate due to absorption of $k$ photons such that $k = \text{int}\left[\frac{I_i}{\hbar \omega_l}\right] + 1$, $I_i$ is the potential of ionization of gas.

In order to construct an equation describing laser beam propagation in nonlinear media let's assume that the beam irradiance distribution, $I(r,z,\eta)$, can be computed from linear Maxwell's equation. In other words we assume that, the beam has undisturbed Gaussian distribution (equation (5)) in the volume where nonlinear refraction occurs (i.e. in the near filed the distortion of laser beam profile is negligible). Of course, in the far field the small changes of the angles of the wave front vector result in substantial changes of beam profile. Such assumption results in significant simplification that, without diminishing generality of the obtained results, provides good accuracy of computations.

Assumption that the distortion of the laser beam irradiance can be neglected during propagation through near focal region is equivalent to the assumption that the *r*-coordinate of an individual ray is controlled mainly by diffraction (i.e. contribution to the change of *r*-coordinate of a ray due to nonlinear effect is neglected). Then, in the reference coordinate system moving with the laser pulse, the angle, $\varphi(r,z_s,\eta)$, of an individual ray propagating from location with coordinate $z=z_0$ to location with coordinate $z=z_s$ can be expressed as a sum of the angle due to diffraction, $\varphi^D(r,z_s)$, given by equation (3), and the angle due to nonlinear refraction, $\varphi^{NLP}(r,z_s,\eta)$, obtained as a result of integration of the equation (8) for a given initial conditions, $\varphi_0(r,z=z_0)$:

$$\varphi(r,z_s,\eta) = \varphi^D(r,z_s) + \varphi^{NLP}(r,z_s,\eta) = \varphi^D(r,z_s) + \int_{z_0}^{z_s} d\varphi^{NLP} = \varphi^D(r,z_s) + \int_{z_0}^{z_s} \frac{\partial n/\partial r}{n} dz \ . \quad (11)$$

### *Laser frequency shift due to self-phase modulation*

Propagation of a laser pulse through a media with refractive index that is dependent on temporally and spatially varying laser irradiance results in laser frequency shift due to self-phase modulation [20]. In the reference coordinate system moving with the laser pulse the shift of the laser frequency while pulse propagates distance *dz* is

$$d_{dz}\omega_l = -\frac{2\pi}{\lambda_0}\frac{dn(z)}{d\eta}dz, \quad (12)$$

where $\lambda_0$ is the laser wavelength in vacuum.

Taking into account equations (9,10) it follows from the equation (12)

$$d_{dz}\omega_l = -\frac{2\pi}{\lambda_0}dz\left[n_2\frac{dI(r,z,\eta)}{d\eta} - \frac{e^2(N_0-N_e)\sigma_k I^k(r,z,\eta)}{2\varepsilon_0 m_e \omega_l^2}\right]. \quad (13)$$

Using common assumption that the temporal profile of the ultrashort laser pulse has Gaussian profile with characteristic duration $\tau_l$, the equation for laser irradiance can be expressed in the following way

$$I(r,z,\eta) = \frac{4E_l}{\pi^{3/2}\tau_l w^2(z)}\exp\left(-\frac{\eta^2}{\tau_l^2}\right)\exp\left(-\frac{2r^2}{w^2(z)}\right), \quad (14)$$

where $E_l$ is laser pulse energy.

Substituting equation (14) into equation (13) and taking integral over the range where nonlinear effect is substantial provides shift of the angular laser frequency, $\Delta\omega_l$,

$$\Delta\omega_l(r,z_s) = \int_{z_0}^{z_s} d_{dz}\omega_l = -\frac{2\pi}{\lambda_0}\int_{z_0}^{z_s} dz\left[n_2\frac{dI(r,z,\eta)}{d\eta} - \frac{e^2(N_0 - N_e)\sigma_k I^k(r,z,\eta)}{2\varepsilon_0 m_e \omega_l^2}\right]. \quad (15)$$

Thus the angular frequency within an individual ray with radial coordinate, $r$, after propagating in the nonlinear medium at the location with $z_s$ coordinate is

$$\omega_l(r,z_s) = \omega_0 + \Delta\omega_l(r,z_s), \quad (16)$$

where $\omega_0$ is the original laser angular frequency.

**Computational procedure**

The propagation equation (11) and the equation (15) for the frequency shift were solved numerically for various time moments, $\eta$, in the coordinate system moving with the laser pulse and for the following initial conditions

$$\varphi_0(r,z=z_0,\eta=\eta_0) = \varphi^D(r,z=z_0) = \arcsin\left(\frac{r}{z_0\left(1+\left(\frac{\pi w_0^2}{\lambda z_0}\right)^2\right)}\right), \quad (17)$$

$$\Delta\omega_l(r,z=z_0,\eta=\eta_0) = 0. \quad (18)$$

The results of computations reported here satisfy two requirements. First, the initial conditions were chosen at a location $z_0<0$ (between the lens and the beam waist) and the time $\eta_0<0$ (before the maximum of the laser pulse power set to be at $\eta_0=0$) for which the laser beam irradiance is small to produce noticeable nonlinear effect (i.e. only diffraction divergence is responsible for the beam evolution). Second, the maximum radial displacement of the individual rays after propagating from the initial location, $z=z_0$, to final location, $z=z_s$, within the computational region does not exceed 5% of the beam radius at the final location.

Verification of the first requirement was performed by integrating the propagation equation (11) from a different initial coordinate (negative value) $z_0$ to a fixed final coordinate $z_s$ (positive value) for various time moments, $\eta$, and determining the smallest value of $z_0$ (most distant in the negative direction from the beam waist) when the output function $\varphi(r,z_s,\eta)$ stop changing noticeably with $z_0$. This value of $z_0$ was used for the initial coordinate and the final coordinate was set at $z_s = -z_0$. It was observed that for all considered conditions the integration range $[z_0; z_s]$ was within $[-2z_R; 2z_R]$, where $z_R$ is the Rayleigh length ($z_R = \frac{\pi w_0^2}{\lambda z_0}$).

The verification of the second requirement was performed by integrating the equation for the radial displacement of an individual ray for various time moments:

$$\Delta r(r) = \int_{z_0}^{z_s}\left(\int_{z_0}^{z_s} d\varphi^{NLP}\right)dz = \int_{z_0}^{z_s}\left(\int_{z_0}^{z} \frac{\partial n/\partial r}{n} dz\right)dz \ . \tag{19}$$

The maximum value of normalized ray radial displacement, $\max\{\Delta r(r)/w(z_s)\}$, was compared with the corresponding dimensionless beam radius $r/w(z_s)$. Only those computations in which the maximum normalized ray displacement was less than 5% of the dimensionless beam radius were accepted.

The computations were performed for a laser wavelength of λ = 800 nm, Gaussian temporal laser pulse shape with characteristic duration $\tau_l$=100fs, and a Gaussian laser beam irradiance radial distribution for two values of beam radius in focal plane $w_0$ = 50 μm and $w_0$ = 200 μm.

We assumed that the main ion produced in air is $O_2^+$. The Keldysh parameter [19], $\gamma = \frac{\omega\sqrt{2mI_i}}{eE}$, was estimated for oxygen molecules $I_i = 12.2$ eV and for all beam intensity values used in the computations. The estimates confirmed that the limit of multiphoton photoeffect is valid for all computational conditions (γ>1). In the computations we used the second order nonlinear refractive index $n_2$ = 5 $10^{-23}$ $m^2/W$ [21] and we assumed eight quanta absorption with $\sigma_8 \approx 1.5 \cdot 10^{-130}$ $m^{16}/s/W^8$ extrapolated from the computed data [22].

**Simulation results**

The results of simulation of laser beam propagation through air (normal conditions) assuming Gaussian temporal shape of the laser pulse with duration τ = 100 fs and a Gaussian spatial beam profile with the beam radius on $1/e^2$ level in the waist, $w_0$ = 50 μm are shown in Figures 1 and 2 as function of dimensionless laser radius, $r/w(z_s)$, and computed for different moments of dimensionless time, η/τ (negative- before the pulse), and two pulse energies of 0.2 mJ and 0.4 mJ. The angle of the wave vector, $\varphi_{z_s}$, and the frequency shift, $\Delta\omega_l$, computed for propagating from $z_0$ = -2$z_R$ to $z_s$ = 2$z_R$ ($z_R$=19.62 mm) are shown in the Figures 1a and 2a and Figures 1b and 2b, correspondingly. Similar data computed for the beam with diameter in the waist, $w_0$ = 200 μm, and the pulse energies 2.5 mJ and 5 mJ are shown in the Figures 3 and 4.

If a laser beam propagated under conditions when medium response is linear, the angle $\varphi_{z_s}$ depends linearly on the dimensionless beam radius, i.e. the wave front is spherical as expected when only the diffraction governs the beam propagation. In nonlinear medium for lower pulse energies when ionization of air is negligible (Figures 1a and 3a) the contribution to the refractive index due to Kerr effect results in focusing of the central part of the beam. The divergence of the outer area of the beam is reduced and for the beam radius larger than approximately $w_l$ the self-induced refraction is negligible and the beam diverges govern by diffraction. The Kerr optical power of self-focusing "lens" is increasing with time reaching its maximum when power is at the maximum value (η=0) and then symmetrically (in time) decreases. Simultaneously with refraction the nonlinear Kerr produces shift of the laser frequency shown in the Figures 1b and 3b. The frequency shift due to Kerr effect is symmetric in time domain, i.e. the red shift prior to the pulse maximum and the blue shift after the pulse maximum have the same amplitude for the symmetric moments of time.

In nonlinear media that is ionized by the laser radiation the contribution of Kerr effect that produces focusing is combined with the defocusing effect due to nonuniform ionization (Figures 2a and 4a). The effect of media ionization on the beam propagation results in strong divergence of the near-axis part of the beam. The maximum angle of divergence is increasing in time as the degree of ionization grows reaching its maximum at the end of the laser pulse. Simultaneously, the outer part of the beam behaves similarly to the described above dynamics dominated by the focusing due Kerr effect. The laser frequency shift dominated by the gas ionization is asymmetric in time domain with substantially larger amplitude of blue shift at maximum of the laser pulse, $\eta = 0$ (Figures 2b and 4b).

**Conclusions**

The presented theoretical analysis reveals complex interplay of refraction and laser frequency shift that is distributed through the space and time domains. Further analysis will produce palpable picture of generating white light and colored conical emission. At this point, it is obvious from our theoretical consideration that self-phase modulation theory adequately describes generation of supercontinuum and colored conical emission. Also, it is evident that white light is produced when substantial gas ionization takes place and some fraction of the defocused and blue-shifted part of the beam originating from the near-axis area overlaps in space and mixes in correct proportion with the focused and red-shifted part of the beam that originates from the outer part. This mixing is at any given moment is "two (several?) color" combination and white light is generated as a result of the integration over time of the frequency shifted "two (several?) color" combination. If one would have detector with femtosecond temporal resolution, a dynamic picture of changing color would be observed instead of white light.

The presented calculations show that some fraction of the blue-shifted part of the beam is strongly defocussed from the near-axis area of the beam, and that should form outer blue color ring of conical emission. Also, it is easy to see that the uncompensated "two color" combinations originating from the area of the beam with larger radial coordinate (where refraction is dominated by the Kerr effect) form inner red-shifted color "rings".

To date, the supercontinuum and conical emission of color rings have been extensively studied experimentally and theoretically. Complex physical models were proposed on the basis of a complete system of Maxwell's equations represented as nonlinear Schrodinger's equation [2,3]. However, to the best of our knowledge, these advanced models failed to produce a comprehendible physical picture of supercontinuum and conical emission generation. In contrast, our comprehensive physical approach provides an adequate description of the physical processes and allows for a practical parametric analysis. As an example of practical value, our approach can be used to test the more advanced models and interpret their results. As most of the current theoretical models, our approach is limited to within paraxial approximation and is inaccurate in simulation of multiple refocusing observed in ultrashort laser pulse propagation.

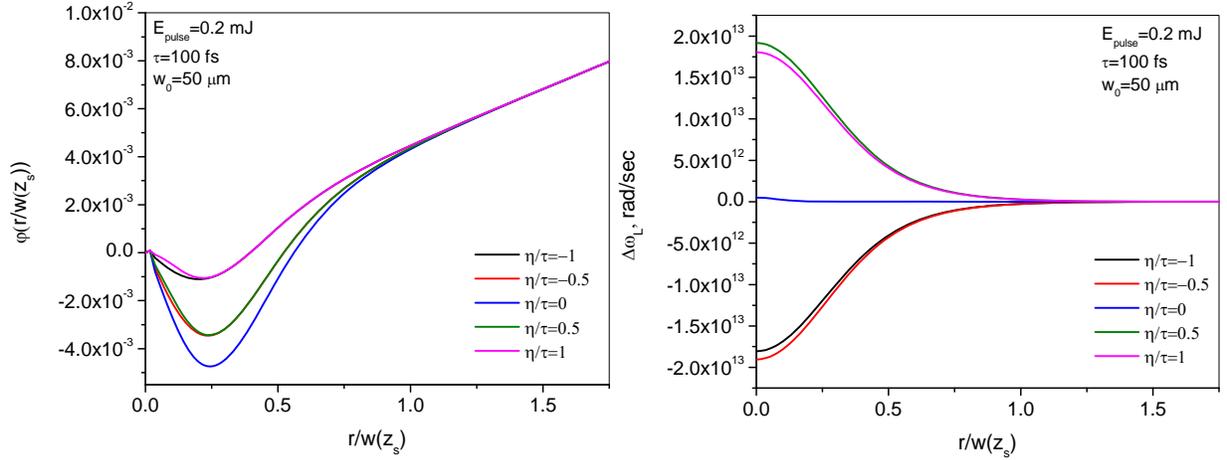

**Fig. 1.** The angle of individual rays (a) and the corresponding frequency changes (b) at a different time moments computed as a function of relative beam radius for a Gaussian (temporally and spatially) beam with characteristic duration of 100 fs and beam radius on $1/e^2$ intensity level of 50μm propagating in air. The total radiated energy per pulse $E_{pulse} = 0.2$ mJ.

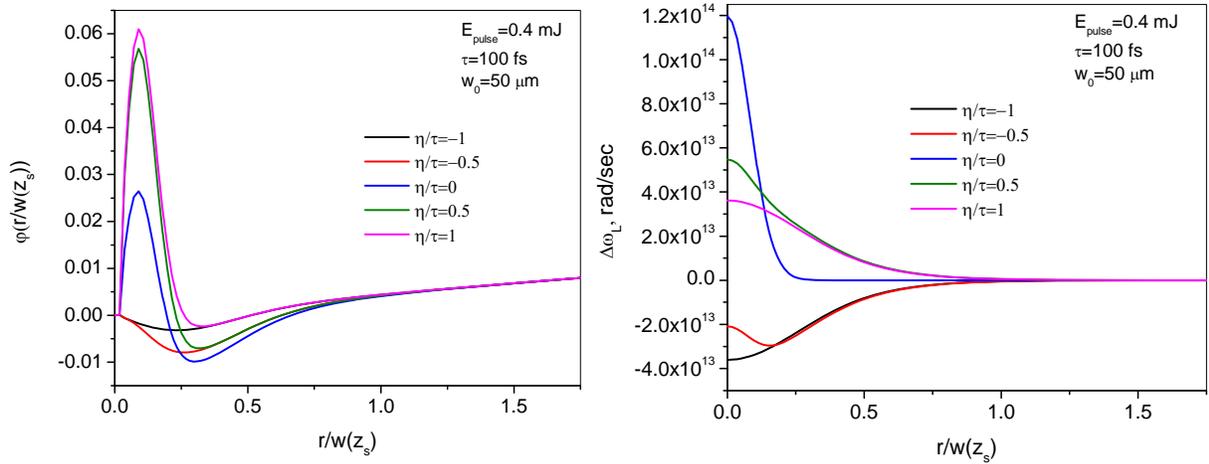

**Fig. 2.** The same as in Fig. 1, but at the total energy per pulse $E_{pulse} = 0.4$ mJ.

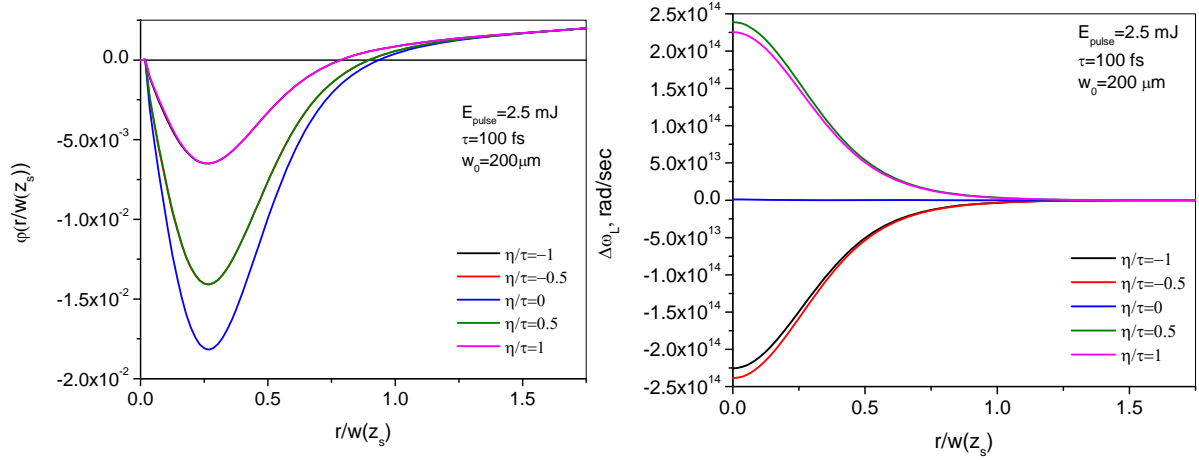

**Fig. 3.** The angle of individual rays (a) and the corresponding frequency changes (b) at a different time moments computed as a function of relative beam radius for a Gaussian (temporally and spatially) beam with characteristic duration of 100 fs and beam radius on $1/e^2$ intensity level of 200μm propagating in air. The total radiated energy per pulse $E_{pulse} = 2.5$ mJ.

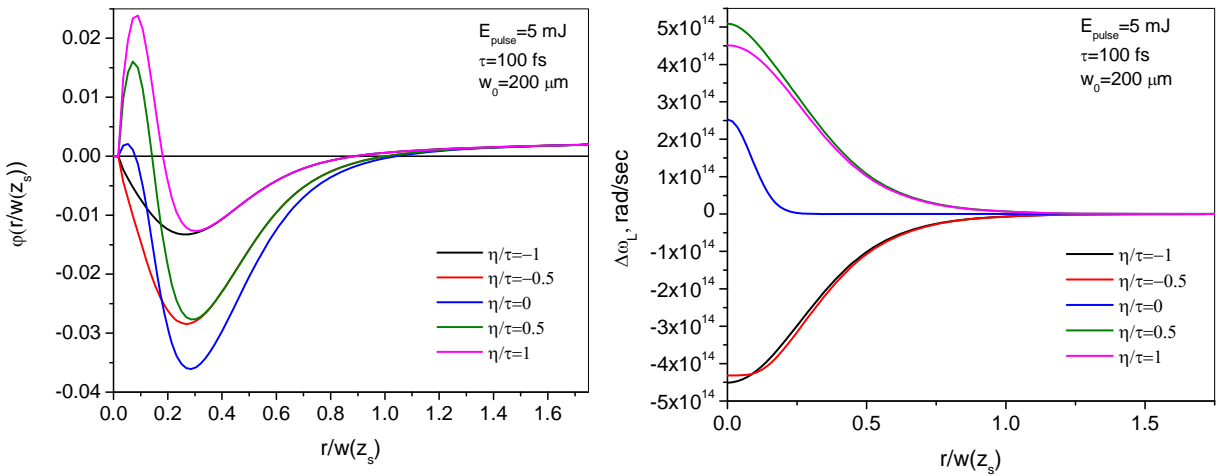

**Fig. 4.** The same as in Fig. 3, but at the total radiated energy per pulse $E_{pulse} = 5$ mJ.